\documentclass{article}

\title{\large Polynomial Time Symmetry and Isomorphism Testing for Connected Graphs}  

\author{Matthew Delacorte}
\date{November 21, 2006}

\begin{document}             

\maketitle 
                  
\footnotesize We use the concept of a Kirchhoff resistor network (alternatively random walk on a network) to probe connected graphs and produce symmetry revealing canonical labelings of the graph(s) nodes and edges.

\section {Voltages}

    \normalsize Electrical networks and random walks on networks have been the object of study for a while (see Doyle,Snell [2]).  Recently these concepts have been taken up by data miners and network analysts to extract useful information from data and networks [3][4][6].  Before this the same techniques where used by computational chemist to label molecules [5].  
    If one takes Kirchhoff's Current Law as a given the exposition of what takes place on a network is easier to follow than using random walks.  Given an undirected connected graph {G = (V,E), {with vertex set $V = \{1,2,...N\}$ and edge set $E = \{1,2,...M\}$ the graph can be represented by an $NxN$ adjacency matrix $A$ where   
 $a_{ij} = w$ if there is an edge of weight $w > 0$ between nodes $i$ and $j$ otherwise $a_{ij} = 0$. 
    Make the edge weights 1 ohm resistors.  Choose one node $a$ to be a current source $I$ and one node $b$ to be a current sink $-I$. Now we apply KCL.
    
\begin{center}
$\sum_j i_{ij}=0$  \quad for $i \neq\ a,b$ \quad $j$ connected to $i$
\end{center}
\begin{center}
$\sum_j i_{aj}=I$
\end{center}
\begin{center}
$\sum_j i_{bj}=-I$
\end{center}

\begin{center}
$i_{ij}=c_{ij}(v_i-v_j)$ \quad $c_{ij}=$ conductance
\end{center} 

\begin{center}
$I\delta(k-a)-I\delta(k-b)=\sum_j c_{kj}(v_k-v_j)$
\end{center}

\begin{center}
$=v_k\sum_j c_{kj}-\sum_j c_{kj}v_j$
\end{center}

For one ohm resistors $c_{ij}=a_{ij}$ \quad $a_{ij}\in A$ where $A$ is the adjacency matrix. \vskip2pt Let $D$ represent the diagonal matrix of vertex degrees $D_{ii}=\sum_j a_{ij}$. \vskip2pt Let $e_i=(0,...,1,...,0_N)$ \vskip2pt Let $I=1$ amp then

\begin{center}
$(e_a-e_b)=(D-A)v$
\end{center}  

\begin{center}
$=Lv$
\end{center} 

Where $L$ is the graph's Laplacian.  $L$ is singular and not invertable.  There is one undetermined degree of freedom. We can solve the KCL equations by using the Laplacian's pseudoinverse $L^{+}$ or we can remove one vertex from the Laplacian and invert the resulting $N-1$ matrix setting the removed vertex voltage to zero.\vskip2pt

$v=L^{+}(e_a-e_b)+\lambda e$ \vskip2pt
 
Where $\lambda e$ is an undetermined constant.\vskip2pt We may also add a universal sink node (an edge to every other node) to the graph to get a $NxN$ ivertable matrix.   Here we can disregard the sink node.

\section {Comparisons}

We now determine the node voltages $v$ over all possible node pairs.  This yields a set of $N$ voltage vectors with $Nx(N-1)$ elements were $Nx(N-1)/2$ of the elements are the negative value of the others.  These voltage vectors may be subtracted from each other to yield a set of $M$ current vectors.  To see if two graphs are isomorphic one does a comparison of the $M$ current vectors of the graphs.  The vector elements may be rearranged at this point, to aid the comparison as their order is immaterial.  To reveal a graph's symetries compare the $N$ node voltage vectors to each other (element order is immaterial) and place them into groups of identical vectors.  These groups of nodes are the orbits of the automorphism group of the graph.

\end{document}